\documentclass[aps,prl,amsmath,amssymb,twocolumn,showpacs]{revtex4-1}
\usepackage[english]{babel}
\usepackage{graphicx}
\usepackage{color}

\newcommand{\abbrev}[1]{\textsc{#1}}

\usepackage[normalem]{ulem}

\begin{document}

\title{Condensate fraction of cold gases in non-uniform external potential}

\author{G.~E.~Astrakharchik$^1$}
\author{K.~V.~Krutitsky$^2$}
\affiliation{$^1$ Departament de F\'{\i}sica i Enginyeria Nuclear, Campus Nord B4-B5, Universitat Polit\`ecnica de Catalunya, E-08034 Barcelona, Spain}
\affiliation{$^2$ Fakult\"at f\"ur Physik der Universit\"at Duisburg-Essen, Campus Duisburg, Lotharstrasse 1, 47048 Duisburg, Germany}

\date{\today}

\begin{abstract}
Exact calculation of the condensate fraction in multi-dimensional inhomogeneous interacting Bose systems which do not possess continuous symmetries is a difficult computational problem. We have developed an iterative procedure which allows to calculate the condensate fraction as well as the corresponding eigenfunction of the one-body density matrix. We successfully validate this procedure in diffusion Monte Carlo simulations of a Bose gas in an optical lattice at zero temperature. We also discuss relation between different criteria used for testing coherence in cold Bose systems, such as fraction of particles that are superfluid, condensed or are in the zero-momentum state.
\end{abstract}

\pacs{67.85.Hj,03.75.Lm}

\maketitle
Bose-Einstein condensation (BEC) is a fascinating phenomenon in which the effects of quantum coherence become apparent on a macroscopic scale~\cite{PS,CBY01,BDZ08}. Although the phenomenon is known for a long time and the basic theoretical concepts are well established, there is no general method which would allow to calculate exactly such fundamental
quantities as the condensate fraction $N_0/N$ and the wave function of the condensate $\phi_0({\bf r})$, which can be probed in modern experiments with ultracold atoms (see, e.g., references in \cite{PS,CBY01,BDZ08}). In order to calculate those, one should solve the eigenvalue problem for one-body density matrix (OBDM) $\rho_1({\bf r};{\bf r}')$~\cite{PO,PS} which is generally a very difficult task. Simple solutions can be obtained for weakly interacting gas at zero temperature. In this case the condensation is almost complete, $N_0/N\approx 1$, and $\phi_0({\bf r})$ can be obtained by means of the mean-field theory as a solution of the Gross-Pitaevskii equation (GPE) $\phi^{GPE}({\bf r})$~\cite{GPE}. Quantum corrections to the mean-field predictions at zero temperature can be calculated within the framework of the Bogoliubov theory~\cite{PS}.

Monte Carlo (MC) methods make no approximation to the model as compared to the perturbative techniques and can address the case of strong interactions. In Ref.~\cite{DuBois} properties of a hard-core Bose gas in a harmonic trap were studied by MC methods. Taking into account spherical symmetry of the confinement and making some additional assumptions, which are valid only for weakly interacting gas, the original problem in three dimensions was reduced to an effective one-dimensional problem and then standard methods of matrix diagonalization were applied to calculate $\phi_0({\bf r})$. However, in a large number of experiments with ultracold atoms the external potentials do not possess spherical symmetry. This is the situation of a gas in the optical lattice~\cite{review,BDZ08}, gas in the presence of a disorder potential~\cite{ReviewDisorder}, {\it etc}. In such cases there is a non-trivial dependence of OBDM on all arguments and it is not possible to reduce the diagonalization problem to a simple matrix formulation. In this Letter we propose a general method for OBDM with arbitrary coordinate dependencies in higher spatial dimensions. We will also compare the condensate fraction with other quantities used for quantitative description of coherence such as the fraction of condensed particles in the momentum space and the superfluid fraction.

The eigenvalue problem for the OBDM reads as
\begin{eqnarray}
\int
\rho_1({\bf r};{\bf r}')
\phi_i({\bf r}')
\,
d{\bf r}'
=
N_i
\phi_i({\bf r})
\;,
\quad
i=0,1,\dots
\label{evp}
\end{eqnarray}
with eigennumbers $N_i$ labeled in descending order and eigenfunctions satisfying the orthonormality condition
\begin{displaymath}
\int
\phi_i^*({\bf r})
\phi_j({\bf r})
\,
d{\bf r}
=
\delta_{ij}
\;.
\end{displaymath}
In order to work out the largest eigenvalue $N_0$ and the corresponding wave function $\phi_0({\bf r})$, we use an idea stemming from the matrix analysis. When a matrix acts on a vector it produces a new vector which can be obtained from the initial one by multiplication of its components along the directions of the eigenvectors by the corresponding eigenvalues. If the resulting vector is renormalized such that it has the same norm as the original one, the result of applying the matrix is a rotation of the vector in the direction of the eigenvector with the largest eigenvalue. Iterating such a rotation many times will eventually align the original vector with the eigenvector with the largest eigenvalue. The convergence of the iterative procedure is very fast if one of the eigenvalues is much larger than the others which is the case of a Bose-condensed system.

Applying this idea to Eq.~(\ref{evp}), we come to the iterative procedure, where the $(i+1)$-th approximation for the wave function $\phi_0^{(i+1)}$ is determined by
\begin{equation}
\int
\rho_1({\bf r};{\bf r}')
\phi_0^{(i)}({\bf r}')
\,
d{\bf r}'
=
N_0^{(i)}
\phi_0^{(i+1)}({\bf r})
\;.
\label{iter}
\end{equation}
The $i$-th approximation for the number of condensed particles is given by
\begin{equation}
N_0^{(i)}
=
\iint
\rho_1({\bf r};{\bf r}')
\phi_0^{(i)*}({\bf r})
\phi_0^{(i)}({\bf r}')
\,
d{\bf r}
\,
d{\bf r}'
\;,
\label{N0i}
\end{equation}
which follows from Eq.~(\ref{evp}). Repeating this procedure permits to obtain, in principle, the condensate wave function exactly. A reasonable choice for the initial approximation is $\phi_0^{(0)}({\bf r})=\phi^{GPE}({\bf r})$.

During the iterations the value of $N_0$ is approached from below. This can be seen by first expanding $\phi_0^{(i)}({\bf r})$ in terms of eigenfunctions of the OBDM $\phi_0^{(i)}({\bf r}) = \sum_{j=0}^\infty c_j \phi_j({\bf r})$ with the normalization condition $\sum_{j=0}^\infty \left|c_j\right|^2 = 1$ and then inserting the resulting expression to Eq.~(\ref{N0i}). This leads to inequality
\begin{eqnarray}
\label{inequality}
N_0^{(i)}
\!\!=\!\!
\sum\limits_{i=0}^\infty
\!
N_i\!\left|c_i\right|^2
\!
\leq
\!
N_0
-
\left(
    N_0\!-\!N_1
\right)\!\!
\left(
    1 \!-\! \left|c_0\right|^2
\right)
\le N_0
\;,
\end{eqnarray}
where $N_1$ is the upper bound for the eigenvalues with $i=1,2,\dots$ This proves the statement.

The iterative procedure can be implemented in MC calculations. The OBDM in first quantized form is expressed in terms of the many-body wave function of the ground state $\psi({\bf R})$ as~\cite{PO,PS}
\begin{eqnarray}
\rho_1({\bf r};{\bf r}')
=
N
\int
\left.
    \psi^*({\bf R})
\right|_{{\bf r}_1={\bf r}}
\left.
    \psi({\bf R})
\right|_{{\bf r}_1={\bf r}'}
d{\bf r}_2
\dots d{\bf r}_N
\;,
\nonumber
\end{eqnarray}
where ${\bf R} = ({\bf r}_1,\dots,{\bf r}_N)$ is a shortcut for a point in $3N$-dimensional phase space. This allows to rewrite Eqs.~(\ref{iter}),~(\ref{N0i}) in a form which can be interpreted in terms of a MC algorithm
\begin{eqnarray}
\frac{N^{(i)}_0}{N}
\phi_0^{(i+1)}\!({\bf r})
\!
=
\!\!\!
\int
\!\!
\left[
\!
\phi_0^{(i)}\!({\bf r}_1)
\frac
{\psi^*({\bf r},{\bf r}_2,\!\dots ,\!{\bf r}_N)}
{\psi^*({\bf R})}
\!
\right]
\!\!
\left|
\psi({\bf R})
\right|^2
d{\bf R}
,
\nonumber
\end{eqnarray}
\begin{eqnarray}
\frac{N^{(i)}_0}{N}
\!\!
=
\!\!\!
\int
\!\!
\left[
\!
\int\!\!
\phi_0^{(i)*}\!({\bf r})
\phi_0^{(i)}\!({\bf r}_1)
\frac{\psi^*\!({\bf r},{\bf r}_2,\dots,{\bf r}_N)}{\psi^*({\bf R})}
d{\bf r}
\right]
\!\!
\left|
\psi({\bf R})
\right|^2
\!
d{\bf R}
.
\nonumber
\end{eqnarray}
MC calculation~\cite{MC} produces a set of configurations ${\bf R}_1,{\bf R}_2,\dots$ in the phase space, distributed according to the best approximation of $\left|\psi({\bf R})\right|^2$. The averaging of each of the quantities in square brackets is then equivalent to the averaging over the produced set of configurations ${\bf R}_1,{\bf R}_2,\dots$. In the case of variational MC method, Metropolis sampling of the trial function $\psi_T({\bf R})$ produces configurations distributed according to $\left|\psi_T({\bf  R})\right|^2$. The diffusion Monte Carlo (DMC) method samples the ``mixed'' distribution $\psi_T({\bf R})\psi({\bf R})$. ``Pure'' averages of $N_0$ and $\phi_0$ over the ground state wave function $\left|\psi({\bf R})\right|^2$ are approximated by extrapolation from variational and mixed estimators. The integral over ${\bf r}$ can be evaluated in a stochastic way by sampling a random point in the simulation box and accumulating the values of the averaged quantity. Since the OBDM is calculated as a mixed estimator, the result that $N_0^{(i)}$ is a lower bound to $N_0$ [Eq.~(\ref{inequality})] should not necessarily hold.

Sufficient condition for the existence of BEC is the off-diagonal long-range order (ODLRO) of the OBDM~\cite{Yang62,BDZ08,CBY01}. If the asymptotic value
\begin{equation}
N_0^{{\bf k}=0}
=
\lim_{|{\bf r}-{\bf r}'|\to\infty}
\rho_1({\bf r};{\bf r}')
\label{ODLRO}
\end{equation}
does not vanish, there is a finite fraction of particles with zero momentum (${\bf k}=0$). $N_0^{{\bf k}=0}$ can be used as a measure of BEC in homogeneous and slightly inhomogeneous systems~\cite{Astrakharchik02a,Pilati}. However, in general the wave function of the state with ${\bf k}=0$ is not the eigenfunction of the OBDM and as it follows from Eq.~(\ref{inequality}) $N_0^{{\bf k}=0}\le N_0$.

Another quantity used to describe coherence in interacting quantum systems is the superfluid fraction. The number of atoms in the superfluid can be obtained as~\cite{sf}
$
N_s
=
\lim_{v \to 0}
2 \Delta F/(m v^2)
$,
where $\Delta F$ is the increase of the free energy in the reference frame moving with the velocity $v$. It is interesting to compare $N_s$ with $N_0$ and $N_0^{{\bf k}=0}$.

In a weak external potential $V({\bf r})$ with the period $L$ in $d$ spatial dimensions, $N_0^{{\bf k}=0}$ can be approximated within perturbative framework as~\cite{BogoliubovGeneralCase}
\begin{equation}
\label{nuc}
\frac{N_0^{{\bf k}=0}}{N}
=
1 -
\sum_{\bf n}
\frac
{\left|\tilde V({\bf n})\right|^2}
{
 L^d\left[\frac{\hbar^2}{2m}
 \left(\frac{2\pi}{L}\right)^2 {\bf n}^2 + 2 g_d n_d\right]^2
}
\;,
\end{equation}
where $g_d$ is an effective interaction parameter in $d$ dimensions, $n_d=N/L^d$ is the number of atoms per unit volume. For a dilute gas with spherically symmetric interaction potential, $g_d$ is proportional to the $s$-wave scattering length $a_s$.
$\tilde V({\bf n})$ is the Fourier transform of the external potential $V({\bf r})$, i.e.,
\begin{eqnarray}
\tilde V({\bf n})
=
\frac{1}{L^{d/2}}
\int\limits_{-L/2}^{L/2}
dr_1
\dots
\int\limits_{-L/2}^{L/2}
dr_d
\;
e^{- i \frac{2\pi}{L} {\bf n} \cdot {\bf r}}
V({\bf r})
\;.
\nonumber
\end{eqnarray}
Equation~(\ref{nuc}) is obtained as a perturbative solution of the GPE~\cite{S-P} and does not take into account quantum fluctuations (Lee-Huang-Yang correction~\cite{LHY}). Analogous calculations for the superfluid fraction lead to an expression similar to Eq.~(\ref{nuc}) but with the coefficient $4/d$ in front of the sum. The same result follows from the Bogoliubov theory. We note that known results for systems with $\delta$-correlated disorder~\cite{disorder} can be reproduced by Eq.~(\ref{nuc}) after statistical averaging.

In order to make a direct comparison between $N_0$, $N_0^{{\bf k}=0}$, and $N_s$, we do numerical simulations of a Bose gas in an optical lattice described by
the following many-body Hamiltonian
\begin{equation}
H
=
\sum_{i=1}^N
\left[
    - \frac{\hbar^2\nabla_i^2}{2m}
    +
    V({\bf r}_i)
\right]
+
\sum_{i<j}
V_{pp}(|\mathbf{r}_i-\mathbf{r}_j|)
\;,
\end{equation}
where $V_{pp}(r)$ is a particle-particle interaction potential.
In the MC calculations, we use the hard-sphere potential of the radius $a_s$.
The GPE is solved for the $\delta$-potential characterized by the scattering length
equal to $a_s$.
The external potential has the form
\begin{equation}
\label{OL}
V({\bf r})
=
V_0
\sum_{\alpha=1}^d
\cos^2
\left(
    2\pi \frac{r_\alpha}{\lambda_L}
\right)
\;,
\end{equation}
where $\lambda_L/2$ is the lattice period. For simplicity we consider a quasi-two dimensional ($d=2$) geometry when the system is subjected to such a tight harmonic oscillator trapping in the third dimension $V(z) = m\omega_{\text{ho}}^2z^2/2$ that the energy of particles is small compared to the energy of the harmonic confinement $\hbar\omega_{\text{ho}}$. This corresponds to recent experiments in anisotropic optical lattices, where the confinement in the third dimension was produced by a periodic potential of large amplitude~\cite{experlattice}. For this setup, Eq.~(\ref{nuc}) reduces to
\begin{equation}
\label{condlat}
\frac{N_0^{{\bf k}=0}}{N}
=
1-
\frac{1}{2}
\left(
    \frac{V_0}{4}
\right)^2
\frac
{d}
{
 \left(
     2 E_R + g_d n_d
 \right)^2
}
\;,
\end{equation}
where $E_R=2\hbar^2\pi^2/(m\lambda_L^2)$ is the recoil energy.

In our calculations, the system parameters are chosen to remain in the superfluid part of the phase diagram. Mean-field theory in the tight-binding approximation gives the following critical value for the superfluid--Mott-insulator transition~\cite{BDZ08}:
\begin{equation}
\label{cp}
2dJ/U_d
=
2 \overline{n} + 1 - 2 \sqrt{\overline{n}\left(\overline{n}+1\right)}
\;,
\end{equation}
where $\overline{n}$ is the number of atoms per lattice site, which must be integer,
$J$ is the tunneling rate, and $U_d$ is the interaction parameter.
For $d=2$, we get an estimate
\begin{equation}
\label{JU}
\frac{J}{U_d}
=
\frac{a_{\text{ho}}}{a_s}
\sqrt{\frac{\lambda_L}{\pi a_{\text{ho}}}}
\left(\!
    \frac{2V_0}{\hbar\omega_{\text{ho}}}\!
\right)^{1/4}
\!\!\!\!\!\!
\exp
\!
\left(\!
    -
    \frac{\lambda_L}{\pi a_{\text{ho}}}
    \sqrt{\frac{2V_0}{\hbar\omega_{\text{ho}}}}
\right)
\;,
\end{equation}
where $a_{\text{ho}} = \sqrt{\hbar/m\omega_{\text{ho}}}$ is the  harmonic oscillator length.  In order to remain in the superfluid regime, the values of $2dJ/U_d$ should be larger than that given by Eq.~(\ref{cp}), which leads to restrictions on the values of $V_0$ and $a_s$.

The one-body part of the variational wave function $\psi_T({\bf r})$ used in MC calculation is obtained by solving GPE for a single lattice period. A plausible approximation for the condensate orbital is
$\phi_0({\bf r})=\phi^{GPE}(x,y)\psi_\text{ho}(z)$,
where $\psi_\text{ho}(z)$ is the ground-state wave function of the harmonic oscillator.
We fix the height of the optical lattice to $V_0 = 0.3\,\hbar\omega_{\text{ho}}$ and restrict ourselves to the case of one atom per lattice period ($\overline{n}=1$). The calculations are carried out for $\lambda_L/a_{\text{ho}}=15$, which corresponds to the experimental setup in Ref.~\cite{experlattice}.

\begin{figure}
\includegraphics[width=0.7\linewidth,angle=-90]{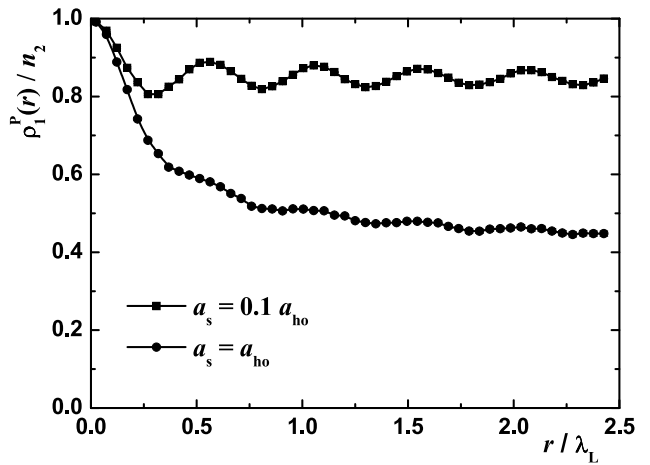}
\caption{
One-body density matrix $\rho^P_1(r)$ normalized to mean density $n_2$ in a system of $N=100$ particles as obtained by extrapolation procedure in DMC calculation. The parameters are $V_0=0.3\,\hbar\omega_{\text{ho}}$, $a_s/a_{\text{ho}} = 0.1$~(squares, upper curve), $1$~(circles, lower curve).
}
\label{Fig1}
\end{figure}

We start the numerical investigation with the case of a weak interaction $a_s/a_{\text{ho}} = 0.1$. According to Eqs.~(\ref{cp}),~(\ref{JU}) the critical value of the lattice strength is  $V_0^c=0.89\;\hbar\omega_{\text{ho}}$.
It turns out from MC calculations that the condensate fraction is very large $N_0/N \approx 0.99$ in a wide range of the strengths of the optical lattice. This means that (i) the system is fully condensed, (ii) the guess $\phi_0({\bf r})=\phi^{GPE}(x,y)\psi_\text{ho}(z)$ for the condensate orbital is indeed extremely good. Figure~\ref{Fig1} shows the averaged OBDM
$
\rho^P_1(r)
=
\int
\frac{d\Omega_{\bf r}}{2\pi}
\int
\frac{d{\bf r}'}{L^2}
\int dz
\rho_1({\bf r}'+{\bf r},z;{\bf r}',z)
$,
where ${\bf r}$ and ${\bf r}'$ are two-dimensional vectors. Its long-range asymptotic value gives the fraction of particles with zero momentum in $(x-y)$ plane. As it is seen from the figure, $N_0^{{\bf k}=0}/N\approx 0.85$ which is considerably smaller than the condensate fraction $N_0/N$. We consider this as a convincing example that the number of particles in the condensate should not be confounded with the number of particles with zero momentum. The GPE approach works very well in the dilute regime and predicts the same fraction of particles with zero momentum, while the result of the Bogoliubov theory is slightly lower $N_0^{{\bf k}=0}/N\approx 0.83$.

The superfluid fraction in the considered case calculated by DMC method $N_s/N = 0.75$ coincides with the value obtained from the GPE. Perturbative Bogoliubov theory gives $N_s/N = 0.67$, refer to Eq.~(\ref{condlat}). It is interesting to note that the superfluid fraction is gradually reduced with increasing $V_0$, while the condensate fraction remains very large. A possible interpretation is that for such small values of the gas parameter the external potential effectively changes very smoothly (i.e. classically), so the system remains well described by the GPE, which corresponds to having almost all particles in the condensate. At the same time the superfluid flow of particles becomes blocked by the strong external field. In this situation, $N_s$ can be much smaller than $N_0$.

Next we study a situation when the condensate fraction is small. To do so, we consider a strongly interacting case with $a_s=a_{\text{ho}}$, for which Eqs.~(\ref{cp}),~(\ref{JU}) predict transition at $V_0^c=0.32\;\hbar\omega_{\text{ho}}$. For this interaction the quantum fluctuations deplete the condensate in a homogeneous system by approximately $20\%$. Presence of a lattice reduces the condensate fraction to $N_0/N = 0.7$. The zero-momentum fraction $N_0^{{\bf k}=0}/N$ is further diminished to about $0.45$ (see Fig.~\ref{Fig1}). GPE as well as Eq.~(\ref{condlat}) give 0.92. Similarly, from the GPE we get $N_s/N = 0.87$, while the Bogoliubov theory predicts a close value $N_s/N = 0.84$. Instead, DMC result $N_s/N = 0.6$ is significantly smaller. This large discrepancy between MC and GPE is not only due to the strong influence of quantum fluctuations but also due to the different forms of the atomic interaction potential.

\begin{figure}
\includegraphics[width=0.7\linewidth,angle=-90]{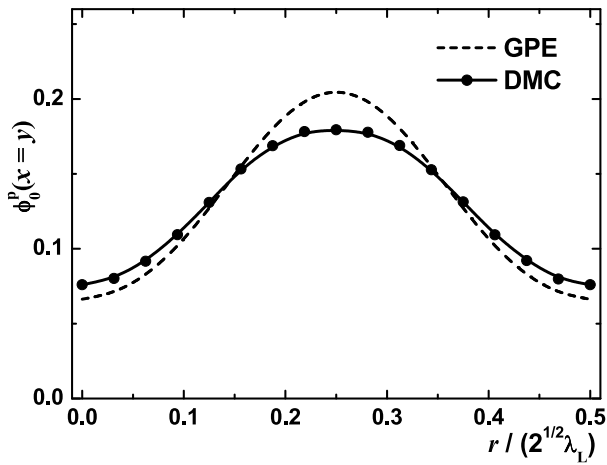}
\caption{Diagonal terms $x=y$ of the projected condensate wave function $\phi_0^p(x,y) = \int \phi_0(x,y,z)\;dz$ obtained as an iteration (\ref{iter}) of the solution of GPE in $(x,y)$ plane multiplied by the Gaussian in $z$ direction.}
\label{Fig2}
\end{figure}

Finally, we test whether the solution of the GPE reproduces well the condensate orbital $\phi_0({\bf r})$. The results are presented in Fig.~\ref{Fig2} for the case of large $s$-wave scattering length $a_s=a_{\text{ho}}$. We find that even in such a strongly interacting system the condensate orbital constructed as a product of $\phi^{GPE}({\bf r})$ by the Gaussian indeed turns out to be almost an eigenstate. On the variational level the Jastrow terms are responsible for suppression of the condensate fraction, although in the studied case the shape of the VMC orbital remains almost unaffected and is very close to the solution of GPE. The DMC algorithm corrects the orbital and makes it less localized compared to the GPE prediction (see Fig.~\ref{Fig2}). Similar effect has been observed in harmonic traps where the condensate moves to the edges\cite{DuBois}.

To conclude, we have developed a procedure of obtaining the number of condensed particles $N_0$ which is applicable to inhomogeneous systems in higher dimensions. For the experimentally relevant case of the Bose gas in the optical lattice, we show that the frequently used criteria of number of particles with zero momentum $N_0^{{\bf k}=0}$ gives a fraction which is smaller than the correct condensate fraction. We propose to approximate the condensate orbital by a solution of GPE and prove that corresponding occupation number is a lower bound to $N_0$. We check in DMC calculations that such an approximation to the condensate orbital can be successfully used even in strongly interacting Bose gases. Numerical results are compared to predictions of perturbative Bogoliubov theory.

Authors are grateful to L.~P.~Pitaevskii for stimulating discussions and ideas that lead to the development of this article.
GEA acknowledges fellowship by MEC (Spain) and financial support by (Spain) Grant No.~\abbrev{fis}2008-04403, Generalitat de Catalunya Grant No.~2009\abbrev{sgr}-1003. The work of KK was supported by the SFB/TR 12 of the German Research Foundation (DFG).


\end{document}